\documentclass{article}
\usepackage{amsmath, amsfonts, amssymb, amsthm, graphicx} 

\newcommand{\cA}{\mathcal{A}}

\newcommand{\cE}{\mathcal{E}}
\newcommand{\cF}{\mathcal{F}}
\newcommand{\cG}{\mathcal{G}}
\newcommand{\cH}{\mathcal{H}}

\newcommand{\cJ}{\mathcal{J}}

\newcommand{\cR}{\mathcal{R}}
\newcommand{\cS}{\mathcal{S}}

\newcommand{\1}{{\mbox{$1$}\rm \!l}}

\newcommand{\wh}{\widehat}
\newcommand{\wY}{{\widehat{Y}}}
\newcommand{\wZ}{{\widehat{Z}}}
\newcommand{\bY}{{\bar{Y}}}
\newcommand{\bZ}{{\bar{Z}}}
\newcommand{\bU}{{\bar{U}}}

\newcommand{\dxi}{{\delta \xi}}
\newcommand{\dZ}{{\delta Z}}
\newcommand{\dY}{{\delta Y}}
\newcommand{\dU}{{\delta U}}

\renewcommand{\theta}{\vartheta}

\newcommand{\Om}{\Omega}
\newcommand{\om}{\omega}

\newcommand{\eps}{\varepsilon}
\newcommand{\lam}{\lambda}

\newcommand{\Lam}{\Lambda}

\newcommand{\dist}{\mathrm{dist}}

\newcommand{\bmo}{\textrm{BMO}}

\newcommand{\IN}{\mathbb{N}}
\newcommand{\IZ}{\mathbb{Z}}

\newcommand{\IR}{\mathbb{R}}
\newcommand{\R}{\mathbb{R}}

\newcommand{\be}{\begin{eqnarray*}}
\newcommand{\ee}{\end{eqnarray*}}
\newcommand{\ben}{\begin{eqnarray}}
\newcommand{\een}{\end{eqnarray}}

\theoremstyle{plain}
\newtheorem{theo}{Theorem}[section]

\newtheorem{lemma}{Lemma}[section]
\newtheorem{propo}{Proposition}[section]
\newtheorem{corollary}{Corollary}[section]

\theoremstyle{definition}

\newtheorem{remark}{Remark}[section]
\newtheorem{example}{Example}[section]

\renewenvironment{proof}[1][] {\smallskip\par\textsl{Proof#1.}\;}{\hspace*{\fill}$\square$\medskip\par}

\begin{document}



\title{CREDIT RISK PREMIA AND QUADRATIC BSDEs WITH A SINGLE JUMP}

%
%
%
%
%

\author{{\vtop{\baselineskip=11pt
\hbox to 125truept{\hss S. ANKIRCHNER\hss}\vskip2truept \hbox
to 125truept{\small\hss Institut f\"ur Angewandte Mathematik \hss}\hbox
to 125truept{\small\hss Rheinische Friedrich-Wilhelms-Universt\"at Bonn \hss}\hbox
to 125truept{\small\hss  Endenicher Allee 60\hss} \hbox to
125truept{\small\hss 53115 Bonn - - Germany\hss}
 \hbox to
125truept{\small\hss ankirchner@hcm.uni-bonn.de\hss}}}\thanks{This research was carried out while S.A.\ visited the \textit{Institut de Science Financi\`ere et d'Assurances (ISFA)} in Lyon for a couple of weeks in the academic year 2007/2008. S.A.\ acknowledges the financial support and the wonderful hospitality received during his visit.}
\and
{\vtop{\baselineskip=11pt \hbox to 125truept{\hss C. BLANCHET-SCALLIET\hss}\vskip2truept \hbox
to 125truept{\small\hss Universit\'e de Lyon\hss} \hbox
to 125truept{\small CNRS, UMR 5208, ICJ\hss}    \hbox
to 125truept{\small\hss Ecole Centrale de Lyon\hss} \hbox
to 125truept{\small\hss Universit\'e Lyon-INSA de Lyon}  \hbox to 125truept{\small\hss 36 avenue Guy de Collongue\hss} \hbox to 125truept{\small\hss 69134 Ecully Cedex - FRANCE\hss}
 \hbox to
125truept{\small\hss christophette.blanchet@ec-lyon.fr \hss} }}
\and
{\vtop{\baselineskip=11pt
\hbox to 125truept{\hss A. EYRAUD-LOISEL\hss}\vskip2truept \hbox
to 125truept{\small\hss Laboratoire SAF \hss}\hbox
to 125truept{\small\hss ISFA - Universit\'e Lyon $1$\hss}\hbox
to 125truept{\small\hss Universit\'e de Lyon\hss} \hbox
to 125truept{\small\hss 50 avenue Tony Garnier\hss} \hbox to
125truept{\small\hss 69007 Lyon - FRANCE\hss}
 \hbox to
125truept{\small\hss anne.eyraud@univ-lyon1.fr\hss} }}
}

\maketitle
%

\begin{abstract}
This paper is concerned with the determination of credit risk premia of defaultable contingent claims by means of indifference valuation principles. Assuming exponential utility preferences we derive representations of indifference premia of credit risk in terms of solutions of Backward Stochastic Differential Equations (BSDE). The class of BSDEs needed for that representation allows for quadratic growth generators and jumps at random times. Since the existence and uniqueness theory for this class of BSDEs has not yet been developed to the required generality, the first part of the paper is devoted to fill that gap.
By using a simple constructive algorithm, and known results on {\em continuous} quadratic BSDEs, we provide sufficient conditions for the existence and uniqueness of quadratic BSDEs with {\em discontinuities} at random times.
\end{abstract}

{\bf AMS Classification (2000):} 60H10,60J75,91B28,91B70,93E20.



\section*{Introduction}\label{sec0}

The current global financial crisis demonstrates the importance of a proper evaluation of credit risk, in particular of financial assets that may default.
In this article we study an approach of determining the credit risk premium of a defaultable contingent claim by using utility indifference principles and techniques of stochastic control. We introduce the concept of an {\em indifference credit risk premium}, defined as the maximal amount of money an owner of a defaultable position is ready to pay for an insurance that completely compensates his credit risk.
We derive a mathematical representation showing that the credit risk premium coincides with the solution of a BSDE.
For an introduction into BSDEs and overview of standard results we refer to \cite{EPQ}.

Credit risk research is a huge field and a large panel of mathematical tools have been used to model, explain and manage credit risk. For an overview we refer the reader to the books of Bielecki and Rutkowski (2002) \cite{BiRu02}, Duffie and Singleton (2003) \cite{DuffieSing03}, Sch\"onbucher (2003) \cite{schon03}, and to the survey papers of Bielecki et al (2004) \cite{BiJbRu04a,BiJbRu04b}.


One standard approach of credit risk modeling is based on the powerful technique of filtration enlargements, by making the distinction between the filtration $\cF$ generated by the continuous processes underlying the market model, and its smallest extension $\cG$ that turns the default time into a $\cG$-stopping time. This kind of filtration enlargement has been referred to as {\em progressive enlargement} of filtrations. It plays an important role in many studies on credit risk, and also the present study strongly profits from this methodology. For an overview on progressive enlargements of filtrations we refer to the fundamental work by Br\'emaud, Yor, Jeulin and the French school of probability in the 80's  \cite{Brem-Yor78}, \cite{Jeu80a}, \cite{Yor78,Yor80}.


For our analysis of credit risk premia we build on the well-known link between BSDEs and the maximal expected utility of economic agents investing in a financial market. We choose the perspective of an economic agent who is investing on an incomplete financial market, and at the same time is holding a contingent claim in her portfolio. For agents with exponential utility preferences and holding {\em non-defaultable} additional claims, it has been shown by Rouge and El Karoui \cite{rouge}, that the maximal expected utility of an economic agent has a representation in terms of a BSDE growing quadratically in the control variable. In \cite{HuImMu05} this result has been generalized to agents that are exposed to non-convex trading constraints.
In this paper, we allow for {\em defaultable} contingent claims in the portfolio, while at the same time extending the tradable assets at the agent's disposal by allowing her to invest (with possibly non-convex trading constraints) in bonds, non-defaultable risky assets and a defaultable zero-coupon bond. We show that in this case the maximal expected utility coincides with the solution of a BSDE that is discontinuous at the default time.


Nearly upon completion of the present article we discovered a related work by Lim and Quenez \cite{LiQu08}, who have also studied the problem of utility maximization of agents endowed with defaultable claims, but who may invest only in a bond and a risky asset. In contrast to our approach using BSDEs as a tool for stochastic control, in \cite{LiQu08} the value function is derived by using dynamical programming techniques. Besides, as a further major difference with our work, it is assumed that the price process of the only risky asset is driven by a Brownian Motion and a default indicating process.

For the BSDEs appearing in the papers \cite{rouge} and \cite{HuImMu05} the authors fall back on existence results of quadratic BSDEs as shown by Kobylanski \cite{koby}. The BSDEs we need for extending their results to defaultable contingent claims, has a discontinuity at the default time, and to our knowledge there are no results in the literature that we may use in order to guarantee the existence of solutions. Therefore, a considerable part of our article is devoted to a thorough analysis of that class of BSDEs. 
More precisely, with $W$ being a multidimensional Brownian motion driving the price processes and $M$ the compensated default process $1_{\{\tau > t\}}$, we consider BSDEs of the form
\ben \label{bsdeintro}
Y_t &=& \xi - \int_t^T Z_s dW_s - \int_t^T U_s dM_s + \int_t^T f(s,Z_s,U_s) ds,
\een
where $\xi$ is the defaultable contingent claim and $f$ is a generator satisfying a quadratic growth condition in $z$.

Notice that quadratic BSDEs similar to \eqref{bsdeintro}, namely with $M$ being replaced by a compensated Levy jump process, have been studied by Morlais \cite{morjump}. Sufficient conditions for the existence of solutions are derived by using approximating BSDEs with generators satisfying a Lipschitz condition.

Having only one possible jump in the BSDE \eqref{bsdeintro}, we propose an alternative approach based on a backward induction in order to derive a solution of \eqref{bsdeintro}. We show that one can find two {\em continuous} quadratic BSDEs from which one can construct a solution of \eqref{bsdeintro} simply by switching from the first to the second one when the default occurs.

We stress that the BSDE \eqref{bsdeintro} has to be solved with respect to the progessively enlarged filtration $\cG$, since the predictable representation property for defaultable claims is valid with respect to $\cG$, but not with respect to the small filtration $\cF$. This bears similarities with initial enlargements. As it is shown in \cite{Anne2005} and \cite{AnneManuela1}, investment decision processes of agents possessing information advantages (represented by an initial enlargement) have to be linked to BSDEs that are solved with respect to enlarged filtration.





\medskip

The paper is organized as follows. In Section \ref{sec1}, we give a precise description of the default model.
Our first aim will be to solve the control problem \eqref{cprb2} by means of BSDE techniques. In Section \ref{sec3} we will show that the maximal expected utility and the optimal strategy can be expressed in terms of a specific BSDE that grows quadratically in the control variable. The BSDE
that will do this job belongs to a class of BSDEs for which no existence theory has been developed yet. In Section \ref{sec2} we make up for that and  clarify the existence and uniqueness of BSDEs with one possible jump. In Section \ref{sec4} we discuss the credit risk premium derived from indifference principles and derive a representation in terms of a BSDE.

\section{The model}\label{sec1}
Let $d\in \IN$ and let $W$ be a $d$-dimensional Brownian motion on a probability space $(\Omega, \cF, P)$. We
denote by $(\cF_t)$ the completion of the filtration generated by $W$.

Our financial market consists in $k$ risky assets and
one non-risky asset. We use the non-risky asset as numeraire and
suppose that the prices of the risky assets in units of the
numeraire evolve according to the SDE
\[ dS^{i}_t = S^{i}_t (\alpha_i(t)  dt + \sigma_i(t) dW_t), \quad i = 1,\ldots, k, \]
where $\alpha_i(t)$ is the $i$th component of a predictable and vector-valued map $\alpha: \Om \times [0,T] \to \R^k$ and $\sigma_i(t)$ is the $i$th row of a predictable and matrix-valued map $\sigma: \Om\times[0,T] \to \R^{k\times d}$.

In order to exclude arbitrage opportunities in the financial market we assume $d \ge k$. Moreover, for technical reasons we suppose that
\begin{enumerate}
\item[(M1)] $\alpha$ is bounded,
\item[(M2)] there exist constants $0 < \eps < K$ such that
$\eps I_k \le \sigma(t) \sigma^*(t) \le K I_k$ for all $t\in[0,T]$,
\end{enumerate}
where $\sigma^*(t)$ is the transpose of $\sigma(t)$, and $I_k$ is the $k$-dimensional unit matrix. Notice that (M1) and (M2) imply that the market price of risk, given by
\be
\theta = \sigma^{*} (\sigma \sigma^{*})^{-1} \alpha,
\ee
is bounded.

\subsection*{Defaultable contingent claims}
We aim at studying pricing and hedging of contingent claims that may default at a random time $\tau: \Omega \to \R_+ \cup \{\infty\}$.
We suppose that at any time, the agent can observe if the default
$\tau$ has appeared or not, which is quite natural to suppose for
default times in finance or for death times in life assurance. So her information is not the
filtration generated by the price processes ${\cal F}$, but is
defined by the following progressive enlargement of filtration, as defined in \cite{blje04} :
\begin{equation}
{\cal G}_t={\cal F}_t\vee \sigma(\1_{\tau\le t}),
\end{equation}
which is the smallest filtration containing the filtration $({\cal
F}_t)$ and that makes $\tau$ a stopping time. Throughout we suppose that the so-called hypothesis ${\bf (H)}$ holds, i.e.
\paragraph{\textbf{Hypothesis}{ \bf
$(H)$}} {\it Any square integrable (${\cal F}, \, P$)-martingale is a square integrable (${\cal G}, \, P$)-martingale}.\\

Under this hypothesis, the Brownian motion $W$ is still a Brownian motion in the enlarged filtration.


Suppose an investor is endowed with a defaultable contingent claim with pay-off $X_1$ at time $T$ if no default occurred, and with a compensation $X_2$ otherwise. Then the total payoff may be written as
\be
F = X_1 1_{\{ \tau > T\}} + X_2 1_{\{ \tau \le T\}}
\ee
where $\tau$ is the default time (or the death time in the case of an
insurance contract). This payoff $F$ consists in a ${\cal
F}_T$-measurable random variable $X_1$, to hedge at maturity $T$ if
$\tau$ has not occurred at time $T$, and a compensation $X_2$,
payed at hit (at the default/death time) in case of default (or
death) occurs before $T$.
In the following we will always assume that $X_1$ is a bounded $\cF_T$-measurable random variable and $X_2=h(\tau)$  where $h$ is a $\cF$-predictable process.\\

Let $D_t = 1_{\{\tau \le t\}}$. Then $D$ is a submartingale, and there exists an $(\cF_t)$-predictable increasing process $K$, called {\em compensator}, such that $K_0 = 0$ and
$$M_t = D_t - \int_0^t (1-D_{s-}) dK_{s}$$ is a martingale with respect to $(\cG_t)$.  We suppose that there exists an $(\cF_t)$-predictable non-negative and bounded process $k_s$ and an $(\cF_t)$-predictable increasing process $A$, with values in $\{0,1\}$, such that
\begin{equation}
\label{intensity}
dK_s = k_s ds + dA_s.
\end{equation}
Since $M$ is a $(\cG_t)$-martingale, it can have no $(\cG_t)$-predictable jumps, and consequently, whenever the predictable process $A$ jumps, $M$ does not jump.
\begin{example} \label{ex1}
Let $k$ a bounded non-negative $(\cF_t)$-predictable process and $\Theta$ an exponentially distributed random variable that is independent of the Brownian Motion $W$. Then the compensator $K$, associated to the stopping time
\be
\tau_1 = \inf \{t, \int_0^t k_sds >\Theta\},
\ee
is given by $K_t = \int_0^t k_s ds$.
\end{example}
\begin{example}
Let $a\in \R_+$, $i \in\{1, \ldots, k\}$, and $\tau_2=\inf\{t \ge 0: S^{i}_t\leq a \textrm{ for one } 1\le i \le k\}$. Notice that $\tau_2$ is an $(\cF_t)$-predictable stopping time. Moreover, let $\tau_1$ be the stopping time from Example \ref{ex1}. The compensator $K$, associated to the stopping time
\be
\tau=\tau_1 \wedge \tau_2,
\ee
is given by $dK_t=k_tdt + d 1_{\{\tau_2 \le t\}}$.
\end{example}
The arbitrage free dynamics of a defaultable zero-coupon bond $\rho_t$ is
\begin{equation}
d\rho_t = \rho_{t-} (a_t dt + c_t dW_t - dM_t)
\end{equation}
where $(a_t,c_t)$ are $ \R\times \R^d$-valued measurable processes. For a derivation of the precise dynamics under the risk neutral measure we refer to \cite{blje04}, and for the dynamics under the historical measure to \cite{theseChristophette}.

By an {\em investment process} we mean any $(\cG_t)$-predictable process $(\lam,\lam^\rho)$, where $\lam =
(\lam^{i})_{1\le i\le k}$ takes values in $\IR^k$ (or a constrained subset, as we will suppose in the next paragraph) such that the
integral process $\int_0^t \lam^{i}_r \frac{dS^{i}_r}{S^{i}_r}$ is
defined for all $i \in\{1, \ldots, k\}$, and $\lam^\rho$ with values in $\IR$ (or a constrained subset) such that $\int_0^t\lam^\rho\frac{d\rho_t}{\rho_{t^-}}$ is defined. We interpret $\lam^{i}$
as the value of the portfolio fraction invested in the $i$th
asset, and $\lam^\rho$ the value of the portfolio invested in the defaultable bond.

Let $p_t = \lam_t \sigma_t$ be the image of $\lam$ with respect to $\sigma$, and $q_t = \lam^\rho \rho_{t^-} $. In what follows we mean by a {\em strategy} the image of any investment process.

Investing according to a strategy $(p,q)$ amounts, at time $t$, to a total trading gain of
\be
G^{p,q}_t = \int_0^t (p_s \theta_s + q_s a_s) ds + \int_0^t (p_s + q_s c_s) d W_s - \int_0^t q_s dM_s.
\ee
Let $\cA$ denote the set of all so-called {\em admissible strategies} $(p,q)$, defined as the integrands satisfying
\begin{equation*}
E\int_0^T |p_s|^2 ds + E\int_0^T |q_s|^2ds<\infty.
\end{equation*}

Throughout let $U$ be the exponential utility function with risk aversion coefficient $\eta > 0$, i.e.
\[ U(x) = - e^{-\eta x}. \]
Consider an investor with preferences described by $U$ investing on the financial market. Moreover, suppose that some constraints are imposed on the investor, in such way that at time $t$ the strategies have to stay within a closed set. We will assume that at any time a strategy $(p_t,q_t)$ belongs to a set $C_t = C^1_t \times C^2_t \subset \R^k\sigma_t \times \R \rho_{t^-}$. In addition, we assume that $(0,0) \in C^1_t \times C^2_t$ for all $t$, and that
\begin{equation}  \label{hypHU}
C^1_t \textrm{ is closed,} \qquad C^2_t \textrm{ is bounded}.
\end{equation}

If the investor has a defaultable position $F$ in her portfolio, then her maximal expected utility is given by
\begin{equation} \label{cprb2}
V^F(v) = \sup \Big \{ EU(v+G^{p,q}_T + F): (p,q) \in \cA, \ (p_s,q_s) \in C^1_s \times C^2_s \textrm{ for all } s \in[0,T] \Big \}.
\end{equation}

In the remainder we will derive a representation of the maximal expected utility in terms of a BSDE driven by the Brownian motion $W$ and the compensated jump process $M$. We will see that $V^F$ is equal to the initial value of a process $(Y_t)_{t\in[0,T]}$ being part of the solution of a BSDE. More precisely, let for all $s\in[0,T]$, $p \in \R^k$, $q \in \R$, $z \in \R^d$ and $u\in\R$,
\be
h(s,p,q,z,u) = - p \theta_s -q a_s + \frac12 \eta |p + q c_s - z|^2 + \frac{1}{\eta} (1-D_{s-})k_s \left[ e^{\eta (u+ q)} - 1 - \eta (u+ q)\right],
\ee
and define
\begin{equation}\label{dominate1}
f(s,z,u) = \min_{(p,q) \in C_s} h(s,p,q,z,u).
\end{equation}
We will show that for bounded and $\cG_T$-measurable $F$ there exists a unique solution of the BSDE
\ben \label{bsde1}
Y_t = F - \int_t^T Z_s dW_s - \int_t^T U_s dM_s + \int_t^T f(s,Z_s,U_s) ds,
\een
and that
\ben \label{u=bsde}
V^F(v) = U(v - Y_0).
\een
There are no results in the literature that guarantee that the BSDE \eqref{bsde1} possesses a solution. Therefore, in the next section we will fill this gap, and we study existence and uniqueness of a class of BSDEs that includes the BSDE \eqref{bsde1}. In Section \ref{sec3} we will come back to the model introduced in this section and prove Equality \eqref{u=bsde}.
%
%
%
%
\section{Quadratic BSDEs with one possible jump} \label{sec2}

Let $(\cJ_t)$ be an arbitrary filtration. We denote by $\cH^2(\cJ_t)$ the set of all $(\cJ_t)$-predictable processes $X_t$ satisfying $E\int_0^T  |X_t|^2 ds < \infty$, and by $\cH^\infty(\cJ_t)$ the set of essentially bounded $(\cJ_t)$-predictable processes. The set of all $(\cJ_t)$-optional processes $X$, for which $E(\sup_{s\in[0,T]} |X_s|^p)$ is finite, will be denoted by $\cR^p(\cJ_t)$, and the set of all bounded $(\cJ_t)$-optional processes by $\cR^\infty(\cJ_t)$. Finally, we write $\cS^\infty(\cJ_t)$ for the set of all $(\cJ_t)$-predictable processes such that $E\int_0^T  |X_s|^2 (1-D_{s-})k_s ds < \infty$.

Throughout this section we will consider BSDEs of the form
\ben \label{bsde}
Y_t &=& \xi - \int_t^T Z_s dW_s - \int_t^T U_s dM_s + \int_t^T f(s,Z_s,U_s) ds
\een
where $\xi$ is a bounded $\cG_T$-measurable random variable, and the generator $f:\Omega \times \R_+ \times \R^d \times \R \to \R$ satisfies the following property:
 {
\begin{itemize}
\item[\textbf{(P1)}] The generator can be decomposed into a sum
\be
f(s,z,u)= [l(s,z)+j(s,u)] (1- D_{s-}) + m(s,z)D_{s-},
\ee
where $l:\Omega \times \R_+ \times \R^d \to \R$, $m:\Omega \times \R_+ \times \R^d \to \R$ and $j:\Omega \times \R_+ \times \R \to \R$ satisfy:
\item $l(\cdot,z)$, $m(\cdot,z)$ and $j(\cdot,u)$ are predictable for all $z \in \R^d$ and $u\in\R$ respectively,
\item $l(\cdot,0), m(\cdot,0)$ and  $j(\cdot,0)$ are bounded, say by $\Lam \in \R_+$,
\item there exists a constant $L \in \R_+$ such that for all $z$ and $z' \in \R^d$
\be
|l(s,z)-l(s,z')| + |m(s,z)-m(s,z')| \leq L (1+|z|+|z'|)|z-z'|,
\ee
\item $j\geq 0$, and $j$ is Lipschitz on $(-K,\infty)$ for every $K \in \IR_+$, with Lipschitz constant $L_j(K)$.
\end{itemize}
} 
In addition, we will sometimes assume that the generator $f$ satisfies also
\begin{itemize}
\item[\textbf{(P2)}] There exists a continuous increasing function $\gamma$ such that for all $s\in [0,T]$ and $u, u' \in [-n,n]$, $n\in \IN$,
\be |j(s,u)-j(s,u')|\leq \gamma(n)\sqrt{k_s} \ |u-u'|, \ee where $k$
is defined by (\ref{intensity}).
\end{itemize}

We first prove a priori estimates for BSDEs of the type (\ref{bsde}), before addressing existence and uniqueness of solutions. We remark that one can show existence by approximating the generator by Lipschitz continuous functions, a method employed in \cite{morjump}. However, as we will see, in this case an approximation is not necessary since one can explicitly construct a solution by combining solutions of non-jump quadratic BSDEs.

\subsection{A priori estimates}

We start showing some technical results which will be used in the proof of existence and uniqueness. We first provide a sufficient condition for the process $\int_0^\cdot Z dW_s$, where $Z$ is the control part of a solution of (\ref{bsde}), to be a $\bmo$ (Bounded Mean Oscillation) martingale.

Recall that a continuous square integrable martingale $(M_t)_{t\in[0,T]}$, with quadratic variation $\langle M,M \rangle$, is a $\bmo$ martingale on $[0,T]$ if and only if there exists a constant $C \in \IR_+$ such that for all stopping times $\tau$ with values in
$[0,T]$ we have
\begin{equation}\label{bmo}
E\left[ \langle M,M \rangle_T  - \langle M,M \rangle_{\tau} \Big|\cG_\tau \right]^\frac12 \le C, \quad \textrm{ a.s.}
\end{equation}
The BMO norm $\|M\|_{\bmo}$ is defined to be the smallest constant $C \in \R_+$ for which (\ref{bmo}) is defined.
In the following lemma we collect some properties of BMO martingales that we
will frequently use.
\begin{lemma}[Properties of BMO martingales]\label{bmoeigen}$\phantom{121}$ \\
Let $M$ be a continuous BMO martingale. Then the following properties are satisfied:
\begin{itemize}
\item[1)] The stochastic exponential of $M$,
\[\cE(M)_T = \exp\{M_T-\frac{1}{2}\langle M, M\rangle_T\}, \]
satisfies $E(\cE(M)_T) = 1$, and thus the measure defined by $dQ = \cE(M)_T dP$
is a probability measure.
\item[2)] The process $\hat{M}= M - \langle M,M \rangle$ is a BMO martingale
relative to the measure $Q$ (see Theorem 3.3 in \cite{kazamaki}).
\item[3)] It is always possible to find a $p>1$ such that $\cE(M) \in L^p$. One can determine such a $p$ with the help of the function
\begin{equation*}
\Psi(x)=\Big\{1+\frac{1}{x^2}\log\frac{2x-1}{2(x-1)}\Big\}^{\frac{1}{2}}-1,
\end{equation*}
defined for all $1<x<\infty$. Notice that $\lim_{x\to1^+} \Psi(x)=\infty$ and $\lim_{x\to\infty} \Psi(x)=0$. It holds true that
if $\lVert M\lVert_{\bmo} \ <\Psi(p)$, then $\cE(M) \in L^p$ (see for
example Theorem 3.1 \cite{kazamaki}).
\end{itemize}
\end{lemma}
\begin{lemma}\label{BMO}(BMO property) Let $\xi$ be a bounded and $\cG_T$-measurable random variable, and $f$ a generator satisfying \textbf{(P1)}. Assume that $(Y,Z,U) \in \cR^\infty(\cG_t) \times \cH^2(\cG_t) \times \cH^\infty(\cG_t)$ is a solution of (\ref{bsde}). Then $\int_0^\cdot Z_s dW_s$ is a BMO-martingale. Moreover, its BMO norm depends only on $\|\sup_{s\in[0,T]}|Y_s|\|_\infty$, $L$, $L_j$,  { $T$ and $\Lam$}.
\end{lemma}
\noindent
\begin{proof}
Let $\kappa$ be an upper bound of the process $|Y|$ and $|U|$.
Ito's formula applied to $e^{aY_t}$, with $a\in\R$, yields
\be
e^{aY_t} &=& e^{a Y_0} + \int_0^t a e^{a Y_{s-}} Z_s d W_s + \int_0^t a e^{a Y_{s-}} U_s d M_s - \int_0^t a e^{a Y_{s-}} f(s,Z_s,U_s) ds \\
&&+ \frac12 \int_0^t a^2 e^{aY_{s-}} |Z_s|^2 ds + \sum_{s\le t} e^{aY_{s-}} (e^{a \Delta Y_s} -1 - a\Delta Y_s) .
\ee
 { Let $g(t,z) = (1-D_{t-}) l(t,z) + D_{t-} m(t,z)$. Then Property {\bf $(P1)$} implies
\begin{eqnarray}
-f(s,Z_s,U_s) &\ge& - |g(s,Z_s) - g(s,0)| - |g(s,0)|- |j(s,U_s) - j(s,0)| - |j(s,0)|\nonumber \\
&\ge& -2L (1 + |Z_s|^2) -L_j(\kappa)|U_s|-2\Lam.\nonumber
\end{eqnarray}
} 
Hence, for $a \ge 0$,
\be
e^{aY_t} &\ge& e^{a Y_0} + \int_0^t a e^{a Y_{s-}} Z_s d W_s + \int_0^t a e^{a Y_{s-}} U_s d M_s - \int_0^t 2 a L e^{a Y_{s-}} ds  \\
& & +  (\frac12 a^2 - 2 a L)\int_0^t  e^{aY_{s-}} |Z_s|^2 ds - \int_0^t a {(L_j(\kappa) \kappa+2\Lam)}e^{a Y_{s-}} ds.
\ee
Taking the conditional expectation yields, for arbitrary stopping times $\tau$ with values in $[0,T]$,
\be
& &(\frac12 a^2 - 2 a L) E\left[\int_\tau^T  e^{aY_{s-}} |Z_s|^2 ds| \cG_\tau\right] \\
&\le& E\left[ e^{aY_T} - e^{aY_\tau}| \cG_\tau\right] + E\left[ \int_\tau^T 2 a L e^{a Y_{s-}} ds | \cG_\tau\right] + \int_0^t a {(L_j(\kappa) \kappa+2\Lam)} e^{a Y_{s-}} ds.\\
\ee
Choose $a = 5L$. Then $\frac12 a^2 - a 2 L = \frac52 L^2$, and
\begin{eqnarray}
\frac52 L^2 e^{- 5L \kappa} E\left[\int_\tau^T  |Z_s|^2 ds| \cG_\tau\right]
\le (2 + 10 L^2 T + 5L T(L_j(\kappa) \kappa+2\Lam)) e^{5L \kappa}, \nonumber
\end{eqnarray}
from where we obtain the result.
\end{proof}
Finally, notice that before the default time, the $Y$ part of a
solution of \eqref{bsde} can only be bounded if the control process
$U$ is bounded (since a jump of height $1$ may occur at any time).
The next lemma puts this intuition into mathematical shape.
\begin{lemma} \label{boundedY} Let $(Y,Z,U)$ be a solution of
\eqref{bsde} such that $\sup_{t \in [0,T]} |Y_t| \le \kappa <
\infty$, $P$-a.s. Then $U$ is bounded by $2 \kappa$, $P\otimes
(1-D_{t-})dK_t$-a.s.
\end{lemma}
\begin{proof} Since $K$ is the compensator of $D$ we have \be E\int_0^T |U_{t-}|
\1_{\{|U_{t-}|
> 2 \kappa\}} (1-D_{t-}) dK_t =E\int_0^T |U_{t-}| \1_{\{|U_{t-}| > 2 \kappa\}}
dD_t. \ee The latter integral vanishes since $\{\int_0^T |U_{t-}|
dD_t > 2 \kappa\} \subset \{\sup_{t \in [0,T]} |Y_t| > \kappa\}$,
and thus we obtain the result.
\end{proof}

We are now able to give the following a priori estimates.

Let $\xi^1$ and $\xi^2$ be two bounded $\cG_T$-measurable random variables, $f^1$ and $f^2$ two generators satisfying properties \textbf{(P1)} and \textbf{(P2)}, and let $(Y^{i},Z^{i},U^{i})\in \cR^\infty(\cG_t) \times \cH^2(\cG_t) \times \cH^\infty(\cG_t)$ be solutions of the BSDEs
\[ Y^{i}_t = \xi^{i} -\int_t^T Z^{i}_s dW_s - \int_t^T U^{i}_s dM_s + \int_t^T f^i(s,Z^{i}_s,U^{i}_s) ds.  \]
Let $\dxi = \xi^1 - \xi^2$, $\delta f_s=f^1(s,Z^1_s,U^1_s)-f^2(s,Z^1_s,U^1_s)$, $\dY = Y^{1}-Y^{2}$, $\dZ = Z^{1}-Z^{2}$ and $\dU = U^{1}-U^{2}$.

\begin{theo} \label{apriori2}
Let $f^1$ and $f^2$ satisfy the properties \textbf{(P1)} and \textbf{(P2)}. There exist constants $q\ge 1$ and $C\in\IR_+$, depending only on $T$, {$\Lam$, $L$, $\|k\|_\infty$, $\|\sup_{s\in[0,T]}|Y_s^1|\|_\infty$, $\|\sup_{s\in[0,T]}|Y_s^2|\|_\infty$ and the functions $L_j$ and $\gamma$}, such that
\be
& &E\left[ \sup_{t\in[0,T]} |\dY_t|^2 + \left(\int_0^T (|\dZ_s|^2 + |\dU_s|^2 (1-D_{s-})\frac{k_s}{2}) ds \right) \right] \\
& &  \qquad \le C \left(E\left[|\dxi|^{2q}+\left(\int_0^T |\delta f_s|ds \right)^{2q} \right]\right)^\frac1q.
\ee
\end{theo}
We will prove Theorem \ref{apriori2} in several steps. First observe that for any $\beta \in\R$, Ito's formula yields
\be
e^{\beta t} \dY^2_t &=& e^{\beta T} \dY_T^2 - 2\int_t^T e^{\beta s} \dY_{s-} \dZ_s dW_s - \int_t^T e^{\beta s} \dU_s 2\dY_{s-} dM_s \\
& & + 2 \int_t^T e^{\beta s} \dY_{s-} [f^1(s,Z^1_s,U^1_s) - f^2(s,Z^2_s,U^2_s)] ds \\
& & - \int_t^T e^{\beta s} (\beta \dY^2_{s-} + |\dZ_s|^2) ds - \sum_{t<s\le T} e^{\beta s} \delta U_s^2 (\Delta M_s)^2,
\ee
and since $\Delta M^2=\Delta M$, this can be further simplified to
\be
e^{\beta t} \dY^2_t &=& e^{\beta T} \dY_T^2 - 2\int_t^T e^{\beta s} \dY_{s-} \dZ_s dW_s - \int_t^T e^{\beta s} \dU_s (2\dY_{s-} +\dU_s) dM_s \\
& & + 2 \int_t^T e^{\beta s} \dY_{s-} [f^1(s,Z^1_s,U^1_s) - f^2(s,Z^2_s,U^2_s)] ds \\
& & - \int_t^T e^{\beta s} (\beta \dY^2_{s-} + |\dZ_s|^2 + {(1-D_{s-})}k_s \dU_s^2) ds.
\ee
Next define two processes
\be
H_s = \frac{f^2(s,Z^1_s, U^1_s) - f^2(s,Z^2_s,U^1_s)}{\delta Z_s} \ \textrm{ and } \ J_s = \frac{f^2(s,Z^2_s, U^1_s) - f^2(s,Z^2_s,U^2_s)}{\delta U_s},
\ee
and note that $f^2(s,Z^1_s,U^1_s) - f^2(s,Z^2_s,U^2_s) = H_s \dZ_s + J_s \dU_s$.
Property \textbf{(P2)} implies $|J_s| \le c \sqrt{k_s} (1-D_{s-})$, for some $c \in \R_+$ that depends on $\|\sup_{s\in[0,T]}|Y^1_s|\|_\infty$ and $\|\sup_{s\in[0,T]}|Y^2_s|\|_\infty$.
Property \textbf{(P1)} implies $|H_s| \le L(1 + |Z^1_s| + |Z^2_s|)$, which, together with Lemma \ref{BMO}, guarantees that $\int_0^\cdot H_s dW_s$ is a BMO martingale with a norm bounded by $L\left\|\int_0^\cdot (1 + |Z^1_s|+ |Z^2_s|)dW_s \right\|_{\bmo}$. By defining $\tilde W_t = W_t - \int_0^t H_u du$, we obtain
\be
e^{\beta t} \dY_t^2 &=& e^{\beta T} \dY_T^2 - 2\int_t^T e^{\beta s} \dY_{s-} \dZ_s d\tilde W_s - \int_t^T e^{\beta s} \dU_s (2\dY_{s-} +\dU_s) dM_s \\
& & + 2\int_t^T e^{\beta s} \dY_{s-} J_s \dU_s ds - \int_t^T e^{\beta s} (\beta \dY^2_{s-} + |\dZ|^2_s + (1-D_{s-})k_s \dU_s^2) ds\\
& & +2\int_t^T e^{\beta s} \dY_{s-} \delta f_s ds.
\ee
Notice that $2 |\dY_{s-} J_s \dU_s| \le 2 |\dY_{s-} c \sqrt{k_s}(1-D_{s-}) \dU_s| \le 2 c^2 |\dY_{s-}|^2 + \frac{k_s}{2} (1-D_{s-}) |\dU_s|^2$. By choosing $\beta = 2 c^2$ we get
\begin{eqnarray}\label{fundeq}
& &e^{\beta t} \dY_t^2 + \int_t^T e^{\beta s} (|\dZ|^2_s + (1-D_{s-})\frac{k_s}{2} \dU_s^2) ds \nonumber \\
&\le& e^{\beta T} \dxi^2 +2\int_t^T e^{\beta s}|\delta f_s| |\delta Y_{s-}| ds  \\ \nonumber
& &- 2\int_t^T e^{\beta s} \dY_{s-} \dZ_s d\tilde W_s - \int_t^T e^{\beta s} \dU_s (2\dY_{s-} +\dU_s) dM_s.
\end{eqnarray}

Based on the previous inequality we will first derive a priori estimates with respect to the auxiliary measure $Q$, defined by $\frac{dQ}{dP} = \cE(H \cdot W)_T$. Note that Girsanov's theorem implies that $\tilde W$ is a $Q$-martingale. Moreover, $M$ remains a martingale with respect to $Q$. To show this recall the well-known fact that $M$ is a martingale with respect to $Q$ if and only if $M_t \cE(H \cdot W)_t$ is a $P$-martingale. The latter is satisfied because $\langle M, \cE(H \cdot W) \rangle = 0$, and hence $d(M_t \cE(H \cdot W)_t) = M_t H_t \cE(H \cdot W)_t dW_t + \cE(H \cdot W)_t dM_t$.

\begin{lemma} \label{apriori1}
For all $p > 1$ there exists a constant $C\in\IR_+$ such that
\be
E^Q\left[ \sup_{t\in[0,T]} |\dY_t|^{2p} + \left(\int_0^T (|\dZ_s|^2 + |\dU_s|^2(1-D_{s-})\frac{k_s}{2}) ds \right)^p \right] \le C E^Q\left(|\dxi|^{2p}+\left(\int_0^T |\delta f_s|ds \right)^{2p}\right).
\ee
\end{lemma}

\begin{proof}
Let $p > 1$. Throughout the proof let $C_1, C_2, \dots$ be
constants depending only on $p$, $T$, $L$, $\gamma$, $\|k\|_\infty$,
$\|\sup_{s\in[0,T]}|Y_s^1|\|_\infty$,$\|\sup_{s\in[0,T]}|Y_s^2|\|_\infty$.
First taking the $\cG_t$-conditional expectation with respect to $Q$
on both sides of Inequality (\ref{fundeq}), and then applying Doob's
$L^p$-inequality implies that
\begin{equation}
E^Q [ \sup_{t\in[0,T]} |\dY_t|^{2p}] \le C_1 E^Q \left( |\dxi|^{2p}+
\left(\int_0^T |\delta f_s| |\dY_s| ds\right)^{p}\right).
\end{equation}
Young's inequality yields
\ben \label{young1}
E^Q \left(\int_0^T |\delta f_s| |\dY_s| ds\right)^{p} &\le& E^Q\left( \sup_{s\in[0,T]} |\dY_s|^p \left(\int_0^T |\delta f_s|  ds\right)^{p} \right) \\
&\le& \frac{1}{2C_1} E^Q \left(\sup_{s\in[0,T]} |\dY_s|^{2p}\right) + C_2 E^Q \left(\int_0^T |\delta f_s|  ds\right)^{2p}, \nonumber
\een
which allows us to deduce
\begin{equation} \label{1.part}
E^Q [ \sup_{t\in[0,T]} |\dY_t|^{2p}] \le C_3 E^Q \left( |\dxi|^{2p}+
\left(\int_0^T |\delta f_s|ds\right)^{2p}\right).
\end{equation}
Besides, it follows from (\ref{fundeq}) that
\be
&  & \left( \int_0^T (|\dZ|^2_s + (1-D_{s-})\frac{k_s}{2} \dU_s^2) ds \right)^p \\
&\le& C_4 \left( |\dxi|^{2p} + \left(\int_0^T |\delta
f_s||\dY_{s-}|ds \right)^p + \left| \int_0^T [2 \dY_{s-} \dZ_s
d\tilde W_s - \dU_s (2\dY_{s-} +\dU_s) dM_s] \right|^p \right) \ee
The Burkholder-Davis-Gundy Inequality yields
\ben\label{BDG} &  &
E^Q \left( \int_0^T (|\dZ_s|^2 + (1-D_{s-}) \frac{k_s}{2} \dU_s^2)
ds \right)^p \\ \nonumber &\le& C_5 E^Q \left(  |\dxi|^{2p} +
\left(\int_0^T |\delta f_s||\dY_{s-}|ds \right)^p\right. \\
\nonumber
&&\left.+ \left| \int_0^T (2 \dY_{s-} |\dZ_s|)^2 ds +
\int_0^T
|\dU_s|^2 (2\dY_{s-} +\dU_s)^2 d[M,M]_s \right|^\frac{p}{2}
\right)\\ \nonumber &\le& C_6 E^Q\left( |\dxi|^{2p}+ \left(\int_0^T
|\delta f_s||\dY_{s-}| ds \right)^p \right. \\ \nonumber
&&\left.+
\left| \int_0^T (\dY_{s-} |\dZ_s|)^2 ds\right|^\frac{p}{2} +
\sum_{0<s\le T} |\dU_s|^p |\dY_{s-} +\dU_s|^p \Delta M_s \right)
\nonumber \een
Using the fact that $(ab) \le 2^{p+1}
C_6 a^2 + \frac{1}{2^{p+1} C_6} b^2$ for all $a, b \in \R_+$, we
get
\ben\label{young}
E^Q\left| \int_0^T (2 \dY_{s-} \dZ_s)^2
ds\right|^\frac{p}{2} &\le& 2^pE^Q\left( \sup_{s\in[0,T]} |\dY_s|^p
\left( \int_0^T \dZ_s^2 ds\right)^\frac{p}{2} \right) \\ \nonumber
&\le& 2^p\left(2^{p+1} C_6 E^Q( \sup_{s\in[0,T]} |\dY_s|^{2p}) +
\frac{1}{2^{p+1} C_6} E^Q \left( \int_0^T \dZ_s^2
ds\right)^p\right). \een Combining (\ref{young}) with (\ref{BDG}),
and using an estimate as in (\ref{young1}), we obtain \be
& & E^Q \left( \int_0^T (|\dZ_s|^2 + (1-D_{s-})\frac{k_s}{2} \dU_s^2) ds \right)^p \\
&\le& C_8 E^Q \left( |\dxi|^{2p} + \left(\int_0^T |\delta f_s|ds
\right)^{2p} + \sup_{s\in[0,T]} |\dY_s|^{2p} + \sum_{0<s\le T}
|\dU_s|^p |\dY_s +\dU_s|^p \Delta M_s \right) \ee Notice that by
Lemma \ref{boundedY}, $|\delta U_s| \le \sup_{t\in[0,T]} 2 |\delta
Y_t|$, $P \otimes (1-D_{s-}) dK_s$-a.s.\ and hence \be
& &E^Q \sum_{0<s\le T} |\dU_s|^p |\dY_{s-} +\dU_s|^p \Delta M_s \\
&=& E^Q \int_0^T |\dU_s|^p |\dY_{s-} +\dU_s|^p dM_s + E^Q  \int_0^T |\dU_s|^p |\dY_{s-} +\dU_s|^p (1-D_{s-})k_s ds \\
&\le& C_9 E^Q  \sup_{t\in[0,T]} |\delta Y_t|^{2p}.
\ee
With the previous inequality, and (\ref{1.part}), the estimate stated in the lemma can be deduced.
\end{proof}

\begin{proof}[ of Theorem \ref{apriori2}]
As in Lemma \ref{apriori1}, let $Q$ be defined by $\frac{dQ}{dP} = \cE(H \cdot W)_T$. Property 2) of Lemma \ref{bmoeigen} guarantees that $\int -H_s d\tilde W_s$ is a BMO martingale w.r.t. $Q$, and hence, by Property 3) of Lemma \ref{bmoeigen}, there exists a constant $p^{\prime}>1$ such that $E^Q\left[\cE(-H \cdot \tilde W)_T^{p^{\prime}}\right] < \infty$.

H\"older's inequality yields, with $p$ being the conjugate exponent of $p^{\prime}$,
\begin{eqnarray}
E \sup_{t\in[0,T]} |\dY_t|^2  &\leq & E^Q\left[\cE(-H \cdot \tilde W)_T \left(\sup_{t\in[0,T]} |\dY_t|^2\right) \right]\\{\nonumber}
&\leq & \left(E^Q\left[\cE(-H \cdot \tilde W)_T^{p^{\prime}}\right]\right)^{1/p^{\prime}}\left(E^Q\left[\sup_{t\in[0,T]} |\dY_t|^{2p} \right]\right)^{1/p}.\\{\nonumber}
\end{eqnarray}
Lemma \ref{apriori1} further implies
\begin{eqnarray}
E\sup_{t\in[0,T]} |\dY_t|^2 
&\leq & C_1 \left(E^Q\left[|\dxi|^{2p}+\left(\int_0^T |\delta f_s|ds \right)^{2p} \right]\right)^{1/p}.\\{\nonumber}
\end{eqnarray}
{As $\int_0^. H_s d W_s$ is a BMO martingale w.r.t. $P$, Property 3)
of Lemma \ref{bmoeigen} implies that there exists a constant
$r^{\prime}>1$ such that $E^P\left[\cE(H \cdot W)_T^{r'}\right] <
\infty$. Then, applying H\"older once again to come back to the
initial measure $P$, we obtain, with $r^{\prime}$ denoting the
conjugate of $r$,  }
\begin{eqnarray}
E\sup_{t\in[0,T]} |\dY_t|^2
&\leq & C_2 \left(E\left[|\dxi|^{2pr}+\left(\int_0^T |\delta f_s|ds \right)^{2pr} \right]\right)^{\frac{1}{pr}}. {\nonumber}
\end{eqnarray}

Using similar arguments, we get
\be
E \left(\int_0^T (|\dZ_s|^2 + |\dU_s|^2 (1-D_{s-})\frac{k_s}{2}) ds \right) \le C_3 \left(E\left[|\dxi|^{2pr}+\left(\int_0^T |\delta f_s|ds \right)^{2pr} \right]\right)^{1/pr},
\ee

and hence the proof is complete.
\end{proof}

\subsection{Existence and Uniqueness of the BSDE}
We now discuss existence and uniqueness of quadratic BSDE with one possible jump. First assume that the terminal condition is a sum of the form $\xi 1_{\{ \tau > T\}} + \zeta 1_{\{ \tau \le T\}}$, where $\xi$ and $\zeta$ are bounded random variables measurable with respect to $\cF_T$. The next result guarantees that there exists a solution of
\ben\label{bsdesimple}
Y_t = \xi 1_{\{ \tau > T\}} + \zeta 1_{\{ \tau \le T\}} - \int_t^T Z_s dW_s - \int_t^T U_s dM_s + \int_t^T f(s,Z_s,U_s) ds.
\een

\begin{theo}(Existence) Let $\xi$ and $\zeta$ be two bounded $\cF_T$-measurable random variables, and let $f$ be a generator satisfying \textbf{(P1)}. Then there exists a solution $(Y,Z,U) \in \cR^\infty(\cG_t) \times \cH^2(\cG_t) \times \cH^\infty(\cG_t)$ of (\ref{bsdesimple}).
\end{theo}
\begin{proof}
In the proof we explicitly construct a solution of (\ref{bsde}) starting from two continuous quadratic BSDEs with terminal conditions $\xi$ and $\zeta$ respectively.

First, by referring to standard existence results as shown in
\cite{koby}, we choose a solution $(\wY, \wZ)\in \cH^\infty(\cF_t)
\times \cH^2(\cF_t)$ of the BSDE \ben \wY_t = \zeta - \int_t^T \wZ_s
dW_s  + \int_t^T  {m(s,\wZ_s)} ds. \een Secondly, we define the
stopping time \be \tau_A = \inf \{t\ge 0: A_t = 1 \}, \ee with the
convention $\inf \emptyset = \infty$, and where $A$ is the
increasing process introduced in Section \ref{sec1}. Observe that
$\tau_A$ is an $(\cF_t)$-predictable stopping time, since $A$ is
$(\cF_t)$-predictable. Since $A$ is part of the
compensator of $D$, it may not jump before $D$, and consequently it
must hold $\tau \le \tau_A$.

Next we consider a BSDE with generator

\be
h(s,y,z)&=& l(s,z)+j(s,(\wY_s - y) 1_{\{\tau_A \ge s\}} )+(\wY_s - y) 1_{\{\tau_A \ge s\}} k_s.
\ee

Since $h$ does not satisfy a Lipschitz condition with respect to $y$, we may not directly invoke standard existence results. However, by using a sandwich argument, we will show that there exist solutions of BSDEs with a bounded terminal condition and generator $h$. For this purpose let
\be
g(s,y,z) = l(s,z)+(\wY_s - y) 1_{\{\tau_A \ge s\}} k_s.
\ee
Let $(Y^g, Z^g)\in \cH^\infty(\cF_t) \times \cH^2(\cF_t)$ be a solution of the BSDE with generator $g$ and terminal condition
\be
\psi = \xi 1_{\{\tau_A > T\}} + \zeta 1_{\{\tau_A \le T\}}.
\ee
In particular, there exists a $K\in \R_+$ such that $\sup_{t\in[0,T]} Y^g_t \ge -K$, a.s.

In addition consider an auxiliary generator $h^{a}$ satisfying, for all $z \in \R^d$ and $s\in[0,T]$,
\be
h^a(s,y,z) = \left\{ \begin{array}{ll} h(s,y,z), & \textrm{ if }  y \in [-K, \infty) \\
g(s,y,z) + j(s,(\wY_s+K) 1_{\{\tau_A \ge s\}})\left[1 - (y + K)
\right] , & \textrm{ else}.
\end{array} \right. \ee Notice that $h^{a}$ is Lipschitz continuous
in $y$, and hence we may again fall back on standard existence
results, such as Theorem 2.3 in \cite{koby}, guaranteeing that there
exists a solution $(Y^a, Z^a) \in \cH^\infty(\cF_t) \times
\cH^2(\cF_t)$ of the BSDE with generator $h^a$ and terminal
condition $\psi$.

We next show that $Y^a_t \ge Y^{g}_t$, a.s. To this end we define the $\R^d$-valued predictable process
\be
\beta^i_s = \frac{l(s,Z_s^a)-l(s,Z_s^g) }{ Z^{a,i}_s - Z^{g,i}_s }, \qquad 1 \le i \le d.
\ee
Observe that \textbf{(P1)} guarantees that there exists a constant $K'\in \R_+$ such that $|\beta_s| \le K'(1 + |Z^a_s| + |Z^g_s|)$, a.s., which implies, together with Lemma \ref{BMO}, that $N_t = \int_0^t \beta_s dW_s$ is a BMO martingale. Hence there exists a probability measure $Q$ on $\cF_T$, with density $\frac{dQ}{dP} = \cE(N)_T$, such that $\tilde W_t = W_t - \int_0^t \beta_s ds$ is a $Q$-Brownian motion. Now observe that
\be
Y^a_t - Y^{g}_t &=& - \int_t^T (Z^a_s - Z^g_s) d \tilde W_s + \int_t^T (Y^a_s - Y^{g}_s )1_{\{\tau_A \ge s\}}k_s ds\\
&+&\int_t^T j(s,(\wY_s-Y^a_s) 1_{\{\tau_A \ge s\}})
1_{Y^a_s>-K}+j(s,(\wY_s+K)\1_{\{\tau_A \ge
s\}}){[1-(Y_s^a+K)]} 1_{Y^a_s \le -K} ds \ee
Notice that the pair
of differences $(Y^a - Y^{g},Z^a - Z^{g})$ solves the linear BSDE
\be y_t = - \int_t^T z_s d\tilde{W}_s + \int_t^T  (\varphi_s +
1_{\{\tau_A \ge s\}}k_s y_s)ds \ee where $\varphi_s =
j(s,(\wY_s-Y^a_s) 1_{\{\tau_A \ge s\}})
1_{Y^a_s>-K}+j(s,(\wY_s+K)\1_{\{\tau_A \ge
s\}}){[1-(Y_s^a+K)]}1_{Y^a_s \le -K}$.
The boundedness of
$j(\cdot,0)$, together with the Lipschitz property of $j$ on compact
sets yield that $\varphi$ is bounded. By using the solution formula
for linear BSDEs (see f.ex.\ Prop.\ 2.2 in \cite{EPQ}), one gets the
representation \be Y^a_t - Y^{g}_t = E^Q\left[\int_t^T
\exp\left(\int_t^s 1_{\{\tau_A \ge u\}}k_u du\right) \varphi_s ds
\Big| {\cal G}_t\right], \ee which shows that $Y^a \ge Y^g$, $Q$-
and $P$-a.s.

Since $Y^g$ is bounded from below by $-K$, we therefore have also $Y^a_t \ge -K$, and this further implies that $(Y^a, Z^a)$ solves the BSDE with generator $h$ and terminal condition $\psi$.

Finally, we have now all at our hands for a solution by setting
\be
Y_t = \left\{ \begin{array}{ll} Y^{a}_t, & (\tau > t) \vee (\tau \le t, \tau = \tau_A),\\
                                                                \wY_t, & (\tau \le t, \tau \lvertneqq \tau_A),
                                                                \end{array} \right.
\ee
\be
Z_t = \left\{ \begin{array}{ll} Z^{a}_t, & (\tau > t) \vee (\tau \le t, \tau = \tau_A),\\
                                                                \wZ_t, & (\tau \le t, \tau \lvertneqq \tau_A),
                                                                \end{array} \right.
\ee
and
\be
U_t = \left\{ \begin{array}{ll} \wY_t - Y^{a}_t, & t \le \tau,\\
                                                                0, & t > \tau.
                                                                \end{array} \right.
\ee
Notice that, on the set $B = \{\tau \le t, \tau \lvertneqq \tau_A\}$, we have
\be
Y_t &=& Y_t - Y_\tau + Y_\tau = (\wY_t - \wY_\tau) + (\wY_\tau - Y^{a}_\tau) + Y^{a}_\tau \\
&=& \int_\tau^t \wZ_s dW_s - \int_\tau^t  {m(s,\wZ_s)} ds + \int_0^t U_s dD_s + Y^{a}_0 + \int_0^\tau Z^{a}_s dW_s - \int_0^\tau h(s, Y^{a}_s,Z^{a}_s)ds \\
&=& Y_0 + \int_0^t Z_s dW_s + \int_0^t U_s dM_s - \int_0^t
f(s,Z_s,U_s) ds. \ee On the complementary set $B^c= \{\tau > t \}
\cup \{\tau \le t, \tau = \tau_A\}$, the martingale $M$ has no jumps
on $[0,t]$ and satisfies $M_t = -\int_0^t (1-D_{s-}) k_s ds$. Hence,
on $B^c$, we have \be
Y_t &=& Y^{a}_t =Y^{a}_0 + \int_0^t Z^{a}_s dW_s - \int_0^t h(s, Y^{a}_s,Z^{a}_s)ds \\
&=& Y^{a}_0+ \int_0^t Z^{a}_s dW_s - \int_0^t f(s,Z^{a}_s, (\wY_s - Y^{a}_s)1_{\{\tau_A \ge s\}} )ds - \int_0^t (\wY_s - Y^{a}_s)1_{\{\tau_A \ge s\}} k_s ds \\
&=& Y^{a}_0+ \int_0^t Z^{a}_s dW_s + \int_0^t f(s,Z^{a}_s,U_s)ds +
\int_0^t U_s dM_s. \ee Finally, the terminal condition $Y_T = \xi
1_{\{\tau > T\}} + \zeta 1_{\{\tau \le T\}}$ is satisfied, since on
$B$ we have, $Y_T = \wY_T = \zeta$, and on $B^c$ we have $Y_T =
Y^{a}_T = \xi 1_{\{\tau_A > T\}} + \zeta 1_{\{ \tau_A \le T\}} = \xi
1_{\{\tau > T\}} + \zeta 1_{\{\tau \le T\}}$.

This shows that $(Y,Z,U)$ is a solution of the BSDE (\ref{bsde}). Moreover, the boundedness of $Y^{a}$ and $\wY$ implies that $Y$ and $U$ are bounded, too.
\end{proof}
We next show that we can still solve BSDE (\ref{bsde}), if we allow the compensation to depend on the default time.

\begin{theo}(Existence)
Let $(\bar\zeta(t))_{0\leq t \le T}$ be a  $(\cF_t)$-predictable bounded process, such that $t\mapsto \bar\zeta(t)$ is almost surely right-continuous on $[0,T]$. Let $f$ satisfy \textbf{(P1)} and \textbf{(P2)}.
Then there exists a solution $(Y,Z,U) \in \cR^\infty(\cG_t) \times \cH^2(\cG_t) \times \cH^\infty(\cG_t)$ of the BSDE
\ben\label{generalcomp}
Y_t = \xi 1_{\{ \tau > T\}} + \bar\zeta(\tau) 1_{\{ \tau \le T\}} - \int_t^T Z_s dW_s - \int_t^T U_s dM_s + \int_t^T f(s,Z_s,U_s) ds.\nonumber
\\
\een

\end{theo}
\begin{proof}
To simplify notation, we assume throughout the proof that the process $A$ is equal to zero, and hence that (\ref{intensity}) simplifies to $dK_s = k_s ds$.

Let $\tau_n$, $n\in \IN$, be the discrete approximation of the default time $\tau$ defined by
\[ \tau_n(\om) = \frac{k}{n} \quad \textrm{ if } \quad  \tau(\om) \in \left]\frac{k-1}{n},\frac{k}{n} \right], k \in \IZ_+.  \]

Observe that $\tau_n$ is a stopping time with respect to the filtration $(\cG_t)$.
For all $\frac{k}{n}<T$,
let $(\wY^{k,n}, \wZ^{k,n})$ be the solution of the BSDE
\be
\wY^{k,n}_t = \bar\zeta(\frac{k}{n}) - \int_t^T \wZ^{k,n}_s dW_s  + \int_t^T f(s,\wZ^{k,n}_s,0) ds.
\ee
Let $(\wY^{T}, \wZ^{T})$ be the solution of the BSDE
\be
\wY^{T}_t = \bar\zeta(T) - \int_t^T \wZ^{T}_s dW_s  + \int_t^T f(s,\wZ^{T}_s,0) ds,
\ee
and let $h^n$ be the family of generators such that $h^n(0,y,z) = 0$ and
\be
h^{n}(s,y,z) = f(s,z,\wY^{k,n}_s - y) \textrm{ if } s \in ]\frac{k-1}{n}, \frac{k}{n}] .
\ee

Let $(y^{n}, z^{n})$ be a solution of the BSDE with terminal condition $\xi$ and generator $h^n$. Moreover, let
\be
Y^n_t = \left\{ \begin{array}{ll} y^{n}_t, & t < \tau,\\
                                                                \wY^{k,n}_t, & t    \ge \tau \textrm{ and } \tau_n= \frac{k}{n}\leq T,\\
                                                                \wY^{T}_t, & t  \ge \tau \textrm{ and } \tau_n= \frac{k}{n}>T,
                                                                \end{array} \right.
\ee
\be
Z^n_t = \left\{ \begin{array}{ll} z^{n}_t, & t \le \tau,\\
                                                                \wZ^{k,n}_t, & t    > \tau \textrm{ and } \tau_n= \frac{k}{n}\leq T,\\
                                                                \wZ^{T}_t, & t  \ge \tau \textrm{ and } \tau_n= \frac{k}{n}>T,
                                                                \end{array} \right.
\ee
and
\be
U^n_t = \left\{ \begin{array}{ll} \wY^{k,n}_t - y^n_t, & t \le \tau, t\in ]\frac{k-1}{n},\frac{k}{n}],\frac{k}{n}\leq T\\
                                                                \wY^{T}_t - y^n_t, & t \le \tau, t\in ]\frac{k-1}{n},\frac{k}{n}],\frac{k}{n}> T\\
                                                                0, & t > \tau.
                                                                \end{array} \right.
\ee
To simplify notations assume that $\tau > 0$ a.s. Then
\ben
\label{equation*}
Y^n_t &=& y^{n}_t 1_{\{t<\tau\}} + \sum_k \wY^{k,n}_t 1_{\{t \ge \tau\}} 1_{\{\tau_n = \frac{k}{n}\leq T \}}+ \wY^{T}_t 1_{\{t \ge \tau\}} 1_{\{\tau_n > T \}}  \nonumber \\
& = & y^{n}_{t\wedge\tau} + \sum_k 1_{\{t \ge \tau\}} 1_{\{\tau_n= \frac{k}{n}\leq T \}} \left[(\wY^{k,n}_\tau - y^{n}_\tau) + (\wY^{k,n}_t - \wY^{k,n}_\tau) \right] \nonumber \\
&&+1_{\{t \ge \tau\}} 1_{\{\tau_n>T \}} \left[(\wY^{T}_\tau - y^{n}_\tau) + (\wY^{T}_t - \wY^{T}_\tau) \right]
\een
Notice that
\be
1_{\{t \ge \tau\}} 1_{\{\tau_n= \frac{k}{n}\leq T \}}(\wY^{k,n}_t - \wY^{k,n}_\tau)&=&1_{\{t \ge \tau\}} 1_{\{\tau_n= \frac{k}{n}\leq T \}}\left(\int_\tau^{t} Z^n_s dW_s - \int_\tau^{t} f(s,Z^n_s,0)ds \right)
\ee
and
\be
1_{\{t \ge \tau\}} 1_{\{\tau_n= \frac{k}{n}\leq T \}}(\wY^{k,n}_\tau - y^{n}_\tau)&=&1_{\{t \ge \tau\}} 1_{\{\tau_n= \frac{k}{n}\leq T \}}\int_0^t U^n_s dD_s.
\ee
Then using the same argument for the third term of \eqref{equation*}, we obtain
\be
Y^n_t&=& y^{n}_0 + \int_0^{t\wedge \tau} z^n_s dW_s - \int_0^{t\wedge\tau} h^n(s,y^n_s,z^n_s)ds + \int_0^t U^n_s dD_s \\
& & + \int_\tau^{t \vee \tau} Z^n_s dW_s - \int_\tau^{t\vee \tau} f(s,Z^n_s,0)ds \\
&=& Y^n_0 + \int_0^t Z^n_s dW_s + \int_0^t U^n_s dM_s - \int_0^t f(s,Z^n_s,U^n_s),
\ee
which shows that $(Y^n,Z^n,U^n)$ is a solution in $\cR^\infty(\cG_t) \times \cH^2(\cG_t) \times \cH^\infty(\cG_t)$ of the BSDE
\ben
Y^n_t &=& \xi 1_{\{ \tau > T\}} + (\bar\zeta(\tau_n) 1_{\{ \tau_n\le T\}}+\bar\zeta(T)1_{\{ \tau_n> T\}})1_{\{ \tau\le T\}} \\& & - \int_t^T Z^n_s dW_s - \int_t^T U^n_s dM_s + \int_t^T f(s,Z^n_s,U^n_s) ds. \nonumber
\een
Now let $1\le n < m$. Then we have $\tau_m \le \tau_n$. Moreover, letting $\kappa$ denote a  bound for the process $|\zeta(t)|$, we obtain for all $q\ge2$
\be
& &E\left[|(\bar\zeta(\tau_m) 1_{\{ \tau_m\le T\}}+\bar\zeta(T)1_{\{ \tau_m> T\}}) 1_{\{ \tau \le T\}}- (\bar\zeta(\tau_n) 1_{\{ \tau_n\le T\}}+\bar\zeta(T)1_{\{ \tau_n> T\}} )1_{\{ \tau \le T\}}|^q \right] \\
&\le&  P(\tau_m \le T < \tau_n)  \kappa^q+E|\bar\zeta(\tau_m) - \bar\zeta(\tau_n)|^q,
\ee
which converges to $0$ as $n,m \to \infty$.\\
Since the random variables $|\zeta(t)|$ and $\xi$ are bounded, results of Kobylanski \cite{koby} and Lemma \ref{BMO} imply that the $\cR^\infty$ norms $\|\sup_{t\in[0,T]} |Y^n_t|\|_\infty$ and the BMO norms of $\int_0^\cdot Z^n_s dW_s$ are uniformly bounded in $n$. Consequently we may deduce from the a priori estimates of Theorem \ref{apriori2} that there exists a $q\ge2$ such that
the sequence $(Y^n, Z^n,U^n)$ is Cauchy in $\cR^q(\cG_t) \times \cH^2(\cG_t) \times \cH^q(\cG_t)$, and hence possesses a limit, say $(Y,Z,U)$, which is easily shown to solve the BSDE (\ref{generalcomp}).
\end{proof}

\begin{theo}(Uniqueness)
Let $\xi$ be a bounded $\cG_T$-measurable random variable and $f$ a generator satisfying \textbf{(P1)} and \textbf{(P2)}, then the BSDE  (\ref{bsde}) has a unique solution in $ \cR^\infty(\cG_t) \times \cH^2(\cG_t) \times \cS^\infty(\cG_t)$.
\end{theo}
\begin{proof}
let $(Y^{i},Z^{i},U^{i})\in \cR^\infty(\cG_t) \times \cH^2(\cG_t) \times \cH^\infty(\cG_t)$ be solutions of the BSDEs
\[ Y^{i}_t = \xi -\int_t^T Z^{i}_s dW_s - \int_t^T U^{i}_s dM_s + \int_t^T f(s,Z^{i}_s,U^{i}_s) ds. \]
Let $\dxi = \xi - \xi=0$, $\delta f_s=0$  $\dY = Y^{1}-Y^{2}$, $\dZ = Z^{1}-Z^{2}$ and $\dU = U^{1}-U^{2}$. Applying Theorem \ref{apriori2}, we obtain for a $q> 1$,
\be
E\left[ \sup_{t\in[0,T]} |\dY_t| ^2+ \left(\int_0^T (|\dZ_s|^2 + |\dU_s|^2(1-D_{s-}) \frac{k_s}{2}) ds \right) \right] \le C E\left[|\dxi|^{2q}+\left(\int_0^T |\delta f_s|ds\right)^{2q} \right] =0.
\ee
\end{proof}
\section{Expected utility and optimal investment in terms of BSDEs}\label{sec3}
For all $s\in[0,T]$, $p \in \R^k$, $q \in \R$, $z \in \R^d$ and $u\in\R$ let
\be
h(s,p,q,z,u) = - p \theta_s -q a_s + \frac12 \eta |p + q c_s - z|^2 + \frac{1}{\eta} (1-D_{s-})k_s \left[ e^{\eta (u+ q)} - 1 - \eta (u+ q)\right],
\ee
and define
\begin{equation}\label{dominate}
f(s,z,u) = \min_{(p,q) \in C_s} h(s,p,q,z,u).
\end{equation}

\begin{remark} If it is impossible to invest in the defaultable zero-coupon, i.e the constraints set is ${(p,q)/ \forall t (p_t,q_t)\in C^1_t\times\{0\}}$, then the generator $f$ satisfies
\ben\label{dominate2} f(s,z,u) = \frac12 \eta
\dist^2(z+\frac{1}{\eta} \theta_s, C^1_s) - \theta_s z -
\frac{|\theta_s|^2 }{2 \eta} + \frac{1}{\eta} (1-D_{s-})k_s \left[
e^{\eta u} - 1 - \eta u\right]. \een In this case hypotheses
\textbf{(P1)} and \textbf{(P2)} are easily seen to be fulfilled.
Moreover, the minimum of $h$ on $C_s\times\{0\}$ is attained at $p =
\Pi_{C_s} (z + \frac{1}{\eta} \theta_s)$.
\end{remark}

\begin{theo}\label{uti=bsde}Let $F = X_1 1_{\{ \tau > T\}} + X_2 1_{\{ \tau \le T\}}$ where $X_1$ is a bounded $\cF_T$-measurable random variable and $X_2=h(\tau)$, where $h$ is a $\cF$-predictable bounded process.
Let $(Y,Z,U) \in \cR^\infty(\cG_t) \times \cH^2(\cG_t) \times \cH^\infty(\cG_t)$ be a solution of the BSDE (\ref{bsde}) with generator defined in (\ref{dominate}) and $\xi=F$. We assume that the generator defined in \eqref{dominate} satisfies properties \textbf{(P1)} and \textbf{(P2)}. Then
\[ V^F(v) = U(v - Y_0), \]
and any predictable process $(\widehat p, \widehat q)$ satisfying $P\otimes \lam$-a.s.
\be
h(s,\widehat p_s, \widehat q_s,Z_s,U_s) = \min_{(p,q) \in C_s} h(s,p,q,Z_s,U_s).
\ee
is an optimal strategy.
\end{theo}
The proof of Theorem \ref{uti=bsde} is based on the following lemma.

\begin{lemma}
$U(v+G^{p, q}_t -Y_t)$ is a local supermartingale for every locally square integrable and $(\cG_t)$-predictable processes $p$ and $q$. If in addition $(p_s, q_s) = \textrm{argmin}_{(u,v) \in C_s} h(s,u,v,Z_s,U_s)$, $P\otimes \lambda$-a.s., then
 $U(v+G^{p,q}_t -Y_t)$ is a local martingale.
\end{lemma}
\begin{proof}
Let $p$ and $q$ be locally square integrable and $(\cG_t)$-predictable processes.

Use the abbreviation $G_t = G^{p,q}_t$ and note that an application of Ito's formula to $U(G_t -Y_t)$ yields
\be
U(v+G_t -Y_t) &=& U(v-Y_0) + \int_0^t U'(v+G_{s-} -Y_{s-}) (p_s + q_s c_s- Z_s) dW_s
\\& & - \int_0^t U'(v+G_{s-} -Y_{s-}) (q_s  +U_s) dM_s
\\& & + \int_0^t U'(v+G_{s-} -Y_{s-}) (p_s \theta_s + q_s a_s + f(s,Z_s, U_s)) ds
\\& & + \frac12 \int_0^t U''(v+G_{s-} -Y_{s-}) |p_s + q_s c_s - Z_s|^2 ds
\\& & + \sum_{0<s \le t} U(v+G_{s-} - Y_{s-}) \left[ e^{-\eta (\Delta G_s -\Delta Y_s)} - 1 + \eta (\Delta G_s -\Delta Y_s)
\right]
\ee
Moreover, we may write
\ben
\lefteqn{U(v+G_t -Y_t) = U(v-Y_0) + \textrm{ local martingale}}
\\& & + \int_0^t U'(v+G_{s-} -Y_{s-}) (f(s,Z_s, U_s) - h(s,p_s,q_s,Z_s, U_s)) ds. \label{BVprocess}%
\een
The monotonicity of $U$ and (\ref{dominate}) implies that the bounded variation process in (\ref{BVprocess}) is decreasing and hence that $U(v+G_t - Y_t)$ is a local supermartingale. If in addition $(p_s, q_s) = \textrm{argmin}_{(u,v) \in C_s} h(s,u,v,Z_s,U_s)$, $P\otimes \lambda$-a.s., then the integrand in (\ref{BVprocess}) vanishes, and therefore in this case $U(v+G^{p,q}_t - Y_t)$ is a local martingale.
\end{proof}

\begin{proof}[ of Theorem \ref{uti=bsde}]
Let $(\widehat p, \widehat q)$ be a $(\cG_t)$-predictable process satisfying $h(s,\widehat p_s, \widehat q_s, z,u) = \min_{(p,q) \in C_s} h(s,p,q,z,u)$, $\lam \otimes P$-a.s.
We first show that $(\widehat p_s, \widehat q_s)$ 
is locally square integrable, and that $((\widehat p + \widehat q c) \cdot W)$ is a BMO martingale.

Since $C^2_t$ is bounded, the process $\widehat q$ is bounded. Moreover we have a representation of $\widehat p$ in terms of $\widehat q$, namely $\widehat p_t = \Pi_{C^1_t}(Z_t + \frac{1}{\eta}\theta_t - \widehat q_t c_t)$, and hence we have
\be
|\widehat p|&\leq&|Z+\frac{1}{\eta}\theta- \widehat q c|+|\widehat p -(Z+\frac{1}{\eta}\theta - \widehat q c)|\\
&\leq& 2|Z|+ 2 |\frac{1}{\eta}\theta|+ 2 |\widehat q c|.
\ee
This immediately yields the square integrability of $p$ and $q$, and with Lemma \ref{BMO}, the BMO property of $((\widehat p + \widehat q c) \cdot W)$. 

According to the previous lemma there exists a sequence of stopping times $\tau_n$ converging to $T$, a.s. such that for all $n\ge1$, the stopped process $U(v+G^{\widehat p, \widehat q}_{t\wedge \tau_n} - Y_{t\wedge \tau_n})$ is a martingale.

Next observe that
\be
U(v+G^{\widehat p, \widehat q}_{t} -Y_{t}) &=& - \cE\left( -\eta \int_0^\cdot \left(\widehat p_s + \widehat q_s c_s - Z_s \right) dW_s \right)_t \\
& & \ \times \ \exp\left( \eta (-v+Y_0) + \eta \int_0^t (\widehat q_s  +U_s)dM_s\right) \\
& & \ \times \ \exp\left(- \int_0^t (e^{\eta (\widehat q_s +U_s)} - 1 - \eta (\widehat q_s +U_s))(1-D_{s-}) k_s ds\right)
\ee

Since $U$ and $Y$ are bounded, this yields that $\{U(v+G^{\widehat p, \widehat q}_{\rho} -Y_{\rho}): \rho $ stopping time with values in $[0,T] \}$ is uniformly integrable. 
Moreover, $\lim_n EU(v+G^{\widehat p, \widehat q}_{t\wedge \tau_n} - Y_{t\wedge \tau_n}) = E U(v+G^{\widehat p, \widehat q}_{t} - Y_{t})$, for all $t\in[0,T]$, from which we deduce, $E U(v+G^{\widehat p, \widehat q}_{T} -Y_{T}) = E U(v -Y_{0})$.

Note that for all $(p,q)\in \cA$ we have
\be
E U(v+G^{p,q}_{T} -Y_{T}) \le E U(v+G^{\widehat p, \widehat q}_{0} -Y_{0}) = E U(v -Y_{0}),
\ee
which shows that $(\widehat p, \widehat q)$ is indeed the optimal strategy. Finally, it follows that $V^F(v) = E U(v -Y_{0})$.
\end{proof}

\section{Credit risk premium}\label{sec4}
In this section we show how the results from the previous sections can be applied in order to obtain probabilistic and analytic expressions for the premium to be paid due to the probable default. To keep things simple we assume that there is no tradable defaultable asset and hence the trading constraints are of the form $C_t=C^1_t \times \{0\}$.

Let $\xi$ be a {\em bounded} $\cF_T$-measurable random variable representing the value of a position if no default occurs.
By {\em indifference credit risk premium} we mean the amount of money $c$ such that an investor is indifferent between holding the non-defaultable security $\xi$, or holding the defaultable security $\xi 1_{\{\tau > T\}}$ and receiving a riskless compensation $c$ at time $0$. To define the indifference credit risk premium in a strict sense, denote by $V^\xi(v)$ and $V^{\xi 1_{ \{\tau > T\} }}(v)$ the maximal expected utility of an investor with initial wealth $v$, and endowment $\xi$ and $\xi 1_{ \{\tau > T\} }$ respectively. Then $c$ is defined as the unique real number satisfying
\be
V^\xi(0) = V^{\xi 1_{ \{\tau > T\} }} (c).
\ee
Since we assume that the preferences are determined by the exponential utility function, $c$ does not depend on the initial wealth of the investor. As for the maximal expect utility, the indifference credit risk premium has a representation in terms of a BSDE, too. To this end let $$g(t,z) = \frac12 \eta \dist^2(z+\frac{1}{\eta} \theta_s, C^1_s) -\theta z - \frac{1}{2 \eta}|\theta|^2$$ and let $(\tilde Y, \tilde Z)$ be the solution of the BSDE
\be
\tilde Y_t = \xi - \int_t^T \tilde Z_s dW_s + \int_t^T g(s,\tilde Z_s)ds.
\ee
Then $V^\xi(0) = U(-\tilde Y_0)$. Analogously, let
\[f(t,z,u) = \frac12 \eta \dist^2(z+\frac{1}{\eta} \theta_s, C^1_s) -\theta z - \frac{1}{2 \eta}|\theta|^2 + \frac{1}{\eta} (1-D_{s-})k_s[e^{\eta u} - 1 - \eta u],\]
and let $(Y, Z, U)$ be the solution of the BSDE
\be
Y_t = \xi 1_{ \{\tau > T\} } - \int_t^T Z_s dW_s - \int_t^T U_s dM_s + \int_t^T f(s,Z_s,U_s)ds.
\ee
Then, by Theorem \ref{uti=bsde}, $V^{\xi 1_{ \{\tau > T\} }} (c) = U(c - Y_0)$.

Next we define the $\R^d$-valued predictable process \be \gamma^i_s
= \frac{\dist^2(Z_s+\frac{1}{\eta} \theta_s, C^1_s)- \dist^2(\tilde
Z_s+\frac{1}{\eta} \theta_s, C^1_s)}{Z^{i}_s - \tilde Z^{i}_s },
\qquad 1 \le i \le d. \ee Notice that we have $|\gamma| \le C(1 +
|Z| + |\tilde Z|)$ for some constant $C \in \R_+$. Therefore the
integral process $\int_0^\cdot \gamma_s dW_s$ is a BMO martingale
and hence we may define the probability measure $\widehat P$ with
density \be \frac{d \widehat P}{dP} = \cE( -
\int_0^\cdot \gamma_s dW_s)_T. \ee Girsanov's theorem implies that
$\widehat W_t = W_t + \int_0^t \gamma_s ds$ is a Brownian motion
with respect to $\widehat P$. We are now ready to establish the
representation of the credit risk premium in terms of a BSDE.
\begin{propo}
The indifference credit risk premium satisfies
\be
c = \bY_0,
\ee
where $(\bY, \bZ, \bar U)$ is the solution of the BSDE
\ben \label{CRPbsde}
\bY_t = \xi 1_{ \{\tau \le T\} } - \int_t^T \bar Z_s d \widehat W_s - \int_t^T \bar U_s dM_s + \int_t^T h(s,\bar Z_s, \bar U) ds,
\een
with generator $h(t,z,u) = -\theta z + \frac{1}{\eta} (1-D_{s-}) k_s[e^{\eta u} - 1 - \eta u]$.
\end{propo}
\begin{proof}
The very definition of the indifference credit risk premium implies
$c = Y_0 - \tilde Y_0$. The differences $\bY = Y- \tilde Y$, $\bZ =
Z - \tilde Z$, $\bU = U$, are easily shown to solve the BSDE
(\ref{CRPbsde}).
\end{proof}
\begin{remark}
Notice that once the default occurred while the time horizon is not attained, the terminal condition in the BSDE \eqref{CRPbsde} is equal to $\xi$, and the compensation is equal to the value of the contingent claim.
\end{remark}

In the following we will derive a lower bound for the credit risk premium given as the expectation of the defaultable security $\xi 1_{ \{\tau \le T\} }$.
\begin{corollary}\label{corodavor}
Let $Q$ be the probability measure defined by $$\frac{dQ}{d \widehat
P} = \cE(\int_0^\cdot \theta_s d \widehat W_s)_T.$$ Then the credit
risk premium is bounded from below by $E^Q[\xi 1_{ \{\tau \le T\}
}]$. In particular, if $\xi$ is positive, then the indifference
credit risk premium is positive, too. Moreover, with vanishing risk
aversion $\eta$, the credit risk premium converges to $E^Q[\xi 1_{
\{\tau \le T\} }]$.
\end{corollary}
\begin{proof}
The solution $(\bY, \bZ, \bar U)$ of (\ref{CRPbsde}) satisfies
\be
\bY_t = \xi 1_{ \{\tau \le T\} } - \int_t^T \bar Z_s d\tilde W_s - \int_t^T \bar U_s dM_s + \int_t^T \frac{1}{\eta} (1-D_{s-}) k_s[e^{\eta \bar U_s} - 1 - \eta \bar U_s] ds,
\ee
where $\tilde W =\widehat W - \int_0^t \theta_s ds$ is a $Q$-Brownian motion. Hence,
\be
\bar Y_t = E^Q\left.\left[\xi 1_{ \{\tau \le T\} } + \int_t^T \frac{1}{\eta} (1-D_{s-}) k_s[e^{\eta \bar U_s} - 1 - \eta \bar U_s] ds \right|\cG_t\right] \ge E^Q\left[\xi 1_{ \{\tau \le T\} } |\cG_t\right].
\ee
Notice that $\lim_{\eta \downarrow 0} \frac{1}{\eta} k_s[e^{\eta u} - 1 - \eta u] = 0$, and since $\bar U$ is bounded, we have that with vanishing risk aversion $\eta$, the credit risk premium converges to $E^Q[\xi 1_{ \{\tau \le T\} }]$.
\end{proof}
\begin{remark}
If the number of uncertainties equals the number of assets (i.e. $k=d$), and if there are no constraints (i.e. $C^1_t = \R^k$), then $\widehat P = P$. Moreover, in this case the measure $Q$ defined in Corollary \ref{corodavor} is the risk-neutral fair measure for pricing non-defaultable derivatives on the assets $S^1, \ldots, S^k$.
\end{remark}

\subsection*{Analytic representation for a defaultable Put Option} 
In this subsection we derive an analytic expression for the credit risk premium of a defaultable put option. To keep things simple we suppose that our financial market consists in only one tradable asset with dynamics evolving according to
\be
dS_t = S_t \alpha dt + S_t \sigma dW_t.
\ee

In addition we assume $k = d = 1$, there are no trading constraints and there does not exist a defaultable asset. So the credit risk is the only source for market incompleteness.
Moreover we suppose the compensator $K$ satisfies $dK_t=k(S_t)dt+dA_t$ where $k$ is for example a positive continuous function.

Let $C \in \R_+$ be the strike of a put option with pay-off function $\psi(x) = (C-x)^+$. We will show that the credit risk premium of $\psi(S_T) 1_{\{\tau > T\}}$ is the initial value of a PDE.

We use the fact that solutions of BSDEs can be represented in terms of solutions of PDEs and vice versa. To this end we characterize in more detail the solution of \eqref{CRPbsde} where we set $\xi = \psi(S_T)$. We first solve the BSDE with driver $(s,z) \mapsto h(s,z,0)$ and non-defaultable derivative $\psi(S_T)$ as terminal condition,
\be
\wh Y_t = \psi(S_T) - \int_t^T \wh Z_s dW_s + \int_t^T h(s,\wh Z_s,0) ds.
\ee
It is known that $\wh Y_t = u(t,S_t)$ where $u$ is the solution of the PDE
\ben\label{bspde}
u_t + \frac12 \sigma^2 x^2 u_{xx}= 0, \quad u(T,x) = \psi(x).
\een
The PDE \eqref{bspde} is the Black-Scholes PDE for put options and the solution is known to satisfy $u(t,x) = C\Phi(-d_2) - x\Phi(-d_1)$ with $d_1= \frac{\ln(\frac{x}{C}) + \frac{\sigma^2}{2} (T-t)}{\sigma \sqrt{T-t}}$ and $d_2=d_1 - \sigma \sqrt{T-t}$.

Up to the default time $\tau$, the solution $(\bar Y, \bar Z)$ of \eqref{CRPbsde} coincides with the solution of the BSDE
\be
Y^{a}_t &=& 0 - \int_t^T Z^{a}_s dW_s + \int_t^T [h(s, Z^{a}_s, u(s,S_s) - Y^{a}_s) + (1-D_{s-})(u(s,S_s) - Y^{a}_s) k(S_s)] ds \\
&=&0 - \int_t^T Z^{a}_s dW_s + \int_t^T [-\theta Z^{a}_s +
(1-D_{s-})\frac{k(S_s)}{\eta}( e^{\eta(u(s,S_s) - Y^{a}_s)} -1)] ds.
\ee Moreover, we have $Y^{a}_t = v(t,S_t)$, where $v$ is the
solution of the PDE \ben\label{crppde} v_t + \frac12 \sigma^2 x^2
v_{xx} + \frac{k(x)}{\eta}(e^{\eta(u - v)} - 1) = 0, \quad v(T,x)
=0. \een Notice that the PDE \eqref{crppde} does not depend on the
drift parameter $\alpha$, which is almost impossible to estimate in
practice. To sum up, we have the following result.
\begin{propo}
Conditionally on $S_t = x$ and $\tau > t$, the credit risk premium at time $t$ of a defaultable put option with strike $C$ and maturity $T>t$ is given by $v(t,x)$, where $v$ is the solution of the PDE \eqref{crppde}.
\end{propo}

\section{Conclusion and final remarks}
In this article we studied a class of BSDEs allowing for a jump at a random time and satisfying a quadratic growth condition, and we provided sufficient conditions for the existence and the uniqueness of solutions.
With this at hand, we have generalized BSDE representations of the maximal expected exponential utility of investors endowed with defaultable contingent claims. Finally, we introduced the notion of {\em indifference credit risk premium} of a defaultable contingent claim, and we derived a representation in terms of a BSDE with quadratic growth generator jumping at the default time.

We remark that one may determine not only the indifference price of credit risk associated with a contingent claim, but the indifference value of the defaultable contingent claim {\em itself}. As for the credit risk premium, the indifference value can be shown to be equal to the difference of two continuous BSDEs with quadratic growth, and hence to a {\em single} BSDE with a jump at the default time $\tau$. By using analogue methods as in Section \ref{sec4}, one can thus  generalize representations of indifference prices as derived in \cite{animre}. Moreover, regularity of continuous quadratic BSDEs, as verified in \cite{animre}, allow to write hedging formulas in terms of derivatives of the indifference value with respect to the market price processes.
One can thus extend delta hedging principles to defaultable contingent claims, by linking the optimal hedge to sensitivities of indifference values.

\textbf{Acknowledgement} The authors thank an anonymous referee for
a very careful reading of the manuscript and many valuable comments.
\bibliography{HedgingDef}
\bibliographystyle{plain}

\end{document}